\documentclass[doublecol]{epl2} 
\title{Coefficient of performance under optimized figure of merit in minimally nonlinear irreversible refrigerator}
\shorttitle{COP under optimized figure of merit in minimally nonlinear irreversible refrigerator} %Insert here a short version of the title if it exceeds 70 characters

\author{Y. Izumida\inst{1,2} \and K. Okuda\inst{3} \and A. Calvo Hern\'andez\inst{4} \and J. M. M. Roco\inst{4}}
\shortauthor{Y. Izumida \etal}

\institute{                    
  \inst{1} Department of Applied Physics, The University of Tokyo, 7-3-1 Hongo, Bunkyo, Tokyo 113-8656, Japan\\
  \inst{2} Department of Physics, The University of Tokyo, 7-3-1 Hongo, Bunkyo, Tokyo 113-0033, Japan\\
  \inst{3} Division of Physics, Hokkaido University, Sapporo 060-0810, Japan\\
  \inst{4} Departamento de F{\'\i}sica Aplicada, and
Instituto Universitario de F\'{\i}sica Fundamental y Matem\'aticas (IUFFyM),
Universidad de Salamanca, 37008 Salamanca, Spain
}
\abstract{
We apply the model of minimally nonlinear irreversible heat engines
developed by Izumida and Okuda [EPL {\bf 97}, 10004 (2012)] to
refrigerators. The model assumes extended Onsager relations
including a new nonlinear term accounting for dissipation effects.
The bounds for the optimized regime under an appropriate figure of
merit and the tight-coupling condition are analyzed and successfully compared with those obtained
previously for low-dissipation Carnot refrigerators in the finite-time
thermodynamics framework. 
Besides, we study the bounds for the nontight-coupling case numerically.
We also introduce a leaky low-dissipation Carnot refrigerator and show that it serves as an example of the minimally nonlinear 
irreversible refrigerator, by calculating its Onsager coefficients explicitly.}

\begin{document}

\maketitle

\section{Introduction}
Nowadays, the optimization of thermal heat devices is receiving a
special attention because of its straightforward relation with the
depletion of energy resources and the concerns of sustainable
development. A number of different performance regimes based on
different figures of merit have been considered with special
emphasis in the analysis of possible universal and unified features.
Among them, the efficiency at maximum power (EMP) $\eta_{\rm maxP}$
for heat engines working along cycles or in steady-state is largely
the issue most studied independently of the thermal device nature
(macroscopic, stochastic or quantum) and/or the model
characteristics~\cite{esposito09, benenti10, norma10, tu12b, wu,
durmayaz, tu08, seifert07, seifert08, seifert11, gaveau11,
esposito10a, esposito10b, abe11}. Even theoretical results have been
faced with an experimental realization for micrometre-sized
stochastic heat engines performing a Stirling cycle~\cite{blickle11}.

For most of the Carnot-like heat engine models analyzed in the
finite-time thermodynamics (FTT) framework~\cite{wu, durmayaz} the
EMP regime allows for valuable and simple expressions of the
optimized efficiency, which under endoreversible assumptions
(\textit{i.e.}, all considered irreversibilities coming from the
couplings between the working system and the external heat
reservoirs through linear heat transfer laws) recover the
paradigmatic Curzon-Ahlborn value~\cite{curzon75} $\eta_{\rm maxP}=
1-\sqrt{\tau}\equiv \eta_{\rm CA}$ using $\tau \equiv T_{\rm c}/T_{\rm h}$, 
where $T_{\rm c}$ and $T_{\rm h}$ denote
the temperature of the cold and hot heat reservoirs, respectively. 
A significant conceptual advance was reported by Esposito {\em et al.}~\cite{esposito10c} 
by considering a low-dissipation Carnot heat engine model. 
In this model the entropy production in the hot (cold) heat
exchange process is assumed to be inversely proportional to the time duration of the process 
(see eqs.~(\ref{eq.low.qh}) and (\ref{eq.low.qc}) for more details). 
Then the reversible regime is approached in the
limit of infinite times and the maximum power regime allows us to
recover the Curzon-Ahlborn value when symmetric dissipation is
considered. This derivation does not require the assumption of any specific
heat transfer law nor the small temperature difference.
Considering extremely asymmetric dissipation limits, these authors
also derived that EMP is bounded from the lower and upper sides as
\begin{eqnarray}
\frac{\eta_{\rm C}}{2}\le \eta_{{\rm maxP}}
\le \frac{\eta_{\rm C}}{2-\eta_{\rm C}},\label{eq.emp.bounds}
\end{eqnarray}
where $\eta_{\rm C}\equiv 1-\tau$ is the Carnot efficiency. The Curzon-Ahlborn value
$\eta_{\rm CA}$ (\textit{i.e.}, the symmetric dissipation
limit) is located between these bounds. Additional studies on
various heat engine models~\cite{yan12,tu12a,tu12c} confirmed and
generalized above results.

Work by de Tom\'as {\em et al.}~\cite{carla12} extended the low
dissipation Carnot heat engine model to refrigerators and besides
proposed a unified figure of merit ($\chi$-criterion described below) focusing the attention in the
common characteristics of every heat energy converter (the working
systems and total cycle time) instead of any specific coupling to
external heat sources which can vary according to a particular arrangement. 
For refrigerators, we denote the heat absorbed by the working system from the cold 
heat reservoir by $Q_{\rm c}$, the heat evacuated from the working system 
to the hot heat reservoir by $Q_{\rm h}$
and the work input to the working system by
$W\equiv Q_{\rm h}-Q_{\rm c}$.
For heat engines, we may inversely choose natural directions of heat transfers and work.
That is, we denote the heat absorbed by the working system from the hot heat reservoir by $Q_{\rm h}$, 
the heat evacuated from the working system to the cold heat reservoir by $Q_{\rm c}$ 
and the work output by $W\equiv Q_{\rm h}-Q_{\rm c}$.
Then we can define this $\chi$-criterion
as the product of the converter efficiency $z$ times the heat
$Q_{\rm in}$ exchanged between the working system and the heat reservoir, 
divided by the time duration of cycle $t_{\rm cycle}$:
\begin{eqnarray}
\chi=\frac{z\, Q_{\rm in}}{t_{\rm cycle}},\label{eq.chi.def}
\end{eqnarray}
where $Q_{\rm in}=Q_{\rm c}$, $z=\varepsilon \equiv Q_{\rm c}/W$ for refrigerators and
$Q_{ \rm in}=Q_{\rm h}$, $z=\eta \equiv W/Q_{\rm h}$ for heat engines. 
It becomes power as $\chi=W/t_{\rm cycle}$ when applied to
heat engines and also allows us to obtain the optimized coefficient of
performance (COP) $\varepsilon_{{\rm max\chi}}$ under symmetric
dissipation conditions when applied to refrigerators~\cite{carla12}:
\begin{eqnarray}
\varepsilon_{{\rm
max\chi}}=\sqrt{1+\varepsilon_{\rm C}}-1\equiv\varepsilon_{\rm
CA},\label{eq.maxchi}
\end{eqnarray}
where $\varepsilon_{\rm C}\equiv \tau/(1-\tau)$ is the Carnot COP.
This result could be considered as the genuine counterpart of the
Curzon-Ahlborn efficiency for refrigerators. It was first obtained
in FTT for Carnot-like refrigerators by Yan and Chen~\cite{yan90}
taking as target function $\varepsilon \dot Q_{\rm c}$, where $\dot
Q_{\rm c}$ is the cooling power of the refrigerator, later and
independently  by Velasco \textit{et al.}~\cite{velasco97a,
velasco97b} using a maximum per-unit-time COP and by Allahverdyan
\textit{et al.}~\cite{allahv10} in the classical limit of a quantum
model with two n-level systems interacting via a pulsed external
field. Very recent results by Wang \textit{et al.}~\cite{tu12d}
generalized the previous results for refrigerators~\cite{carla12} and they obtained
the following bounds of the COP considering extremely asymmetric dissipation limits:
\begin{eqnarray}
0\le \varepsilon_{{\rm max\chi}} \le
\frac{\sqrt{9+8\varepsilon_{\rm C}}-3}{2}.\label{eq.cop.bounds}
\end{eqnarray}
These bounds for refrigerators play the same role that eq.~(\ref{eq.emp.bounds}) for heat engines.

A shortcoming of the all above results is the model dependence.
Thus additional research work~\cite{broeck05, borja06, borja07,borja08} 
has been devoted to obtain model-independent results on efficiency and COP using the well
founded formalism of linear irreversible thermodynamics (LIT) for
both cyclic and steady-state models including explicit calculations
of the Onsager coefficients using molecular kinetic
theory~\cite{yuki09, yuki10, yuki08}. Beyond the linear regime a
further improvement was reported by Izumida and Okuda~\cite{yuki12}
by proposing a model of minimally nonlinear irreversible heat
engines described by extended Onsager relations with nonlinear
terms accounting for power dissipation. The validity of the theory
was checked by comparing the results for EMP and its bounds with
eq.~(\ref{eq.emp.bounds}). 
Now, the main goal of the present paper is
to extend the application of this nonlinear irreversible
theory~\cite{yuki12} to refrigerators and analyze their COP under maximum $\chi$-condition. 
To get this we first present the model and analyze the results when the
tight-coupling condition is met. Then we also study the nontight-coupling case numerically.
Finally we extend the low-dissipation Carnot refrigerator model~\cite{tu12d} to a leaky model 
and show that it serves as an example of the minimally nonlinear irreversible refrigerator
by calculating its Onsager coefficients explicitly.
\section{Minimally nonlinear irreversible refrigerator model}
Refrigerators are generally classified into two types, that is, steady-state refrigerators and
cyclic ones. Our theory can be applied to both cases.
Since the working system stays unchanged for steady-state refrigerators and comes back to the original state after one-cycle for cyclic refrigerators,
the entropy production rate $\dot{\sigma}$ of the refrigerators agrees with
the entropy increase rate of the heat reservoirs as
\begin{eqnarray}
\dot{\sigma}=\frac{\dot{Q}_{\rm c}+\dot{W}}{T_{\rm h}}-\frac{\dot{Q}_{\rm c}}{T_{\rm c}}
=\frac{\dot{W}}{T_{\rm h}}+\dot{Q}_{\rm c} \left(\frac{1}{T_{\rm h}}-\frac{1}{T_{\rm c}}\right),\label{eq.entropy-refrigerator}
\end{eqnarray}
where the dot denotes the quantity per unit time for steady-state refrigerators and
the quantity divided by the time duration of cycle $t_{\rm cycle}$ for cyclic refrigerators.
From the decomposition of $\dot{\sigma}$ into the sum of the product of thermodynamic flux $J_i$ and its conjugate thermodynamic force $X_i$ as $\dot{\sigma}=J_1X_1+J_2X_2$, 
we can define $J_1\equiv \dot{x}$, $X_1\equiv F/T_{\rm h}$ and $J_2\equiv \dot{Q}_{\rm c}$,
$X_2\equiv 1/T_{\rm h}-1/T_{\rm c}$ for steady-state refrigerators, where
$\dot{W}=F\dot{x}$ with $F$ and $x$ denoting the time-independent
external generalized force and its conjugate variable, respectively.
Likewise we can also define 
\begin{eqnarray}
J_1\equiv \frac{1}{t_{\rm cycle}},\ X_1\equiv \frac{W}{T_{\rm h}},\label{eq.def_J1_X1}\\
J_2\equiv \dot{Q}_{\rm c}=\frac{Q_{\rm c}}{t_{\rm cycle}},\  X_2\equiv \frac{1}{T_{\rm h}}-\frac{1}{T_{\rm c}},\label{eq.def_J2_X2}
\end{eqnarray}
for cyclic refrigerators, where $X_1$ and $X_2$ play the
role of driver and driven forces, respectively.~\footnote{To establish above
election of the thermodynamic fluxes and forces from eq.~(\ref{eq.entropy-refrigerator}) we mainly
focus on the specific function of the refrigerator system (the
extracted cooling power $\dot Q_{\rm c}$ of the low-temperature reservoir)
as a consequence of the input of an external power $\dot W$~\cite{borja08}. An
alternative starting point is to express the entropy production rate
in eq.~(\ref{eq.entropy-refrigerator}) in terms of $\dot W$ and $\dot Q_{\rm h}$. If this is done the
thermodynamic fluxes and forces are exactly the same as those
obtained for a heat engine in~\cite{yuki12} with the sole difference
coming from the sign convention: $\dot{W}$ is given in (delivered to) for
a refrigerator (heat engine) while $\dot Q_{\rm h}$ is delivered to (given
in) for a refrigerator (heat engine). We have checked that this
alternative election of fluxes and forces does not change the main results in the present paper.}
\begin{figure}
\begin{center}
\includegraphics[scale=0.15]{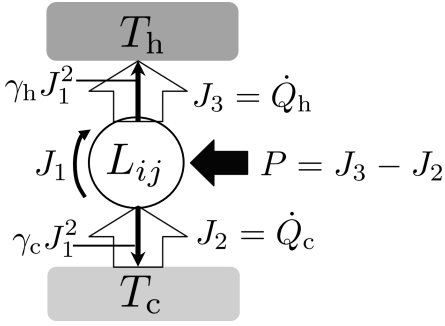}
\end{center}
\caption{Schematic illustration of the minimally nonlinear irreversible
refrigerator described by eqs.~(\ref{eq.J2.by.J1}) and (\ref{eq.J3.by.J1}).}\label{fig.1}
\end{figure}
Anyway, we assume that the refrigerators are described by the following extended Onsager relations~\cite{yuki12}:
\begin{eqnarray}
&&J_1=L_{11}X_1+L_{12}X_2,\label{eq.minimal.J1} \\
&&J_2=L_{21}X_1+L_{22}X_2-\gamma_{\rm c} {J_1}^2, \label{eq.minimal.J2}
\end{eqnarray}
where $L_{ij}$'s are the Onsager coefficients with reciprocity $L_{12}=L_{21}$.
The nonlinear term $-\gamma_{\rm c} J_1^2$ means the power dissipation into the cold
heat reservoir and $\gamma_{\rm c}>0$ denotes its strength. We also disregard
other possible nonlinear terms by assuming that their coefficients are too small 
when compared with the power dissipation term.

In the limit of $X_i$'s $\to 0$, 
the thermodynamic forces can be approximated as 
$X_1\simeq W/T$ ($X_1=F/T$) and $X_2\simeq \Delta T/T^2$ for the cyclic (steady-state) 
refrigerators, where $\Delta T \equiv T_{\rm h}-T_{\rm c}$ and $T\equiv (T_{\rm h}+T_{\rm c})/2$.
These expressions of the forces are consistent with~\cite{broeck05,yuki09}.
Then the extended Onsager relations in eqs.~(\ref{eq.minimal.J1}) and (\ref{eq.minimal.J2}) 
recover the usual linear Onsager relations such as in~\cite{broeck05,yuki09}.
Thus our minimally nonlinear irreversible model naturally includes the linear irreversible model
as a special case.
The non-negativity of the entropy production rate $\dot{\sigma}=J_1 X_1+J_2X_2$, 
which is a quadratic form of $X_i$'s in the linear irreversible model, leads to the restriction of the
Onsager coefficients~\cite{yuki09,O,GM} to
\begin{eqnarray}
L_{11}\ge 0, L_{22}\ge 0, L_{11}L_{22}-L_{12}L_{21} \ge 0.\label{eq.onsager.coeffi.restrict}
\end{eqnarray}
Although our model includes the power dissipation as the nonlinear
term, we also assume eq.~(\ref{eq.onsager.coeffi.restrict}) holds.
This idea that the dissipation is included in the scope of the
linear irreversible thermodynamics is also found in~\cite{CW,AOGGLa,AOGGLb}.

The power input $P$ is given by $P\equiv \dot{W}=J_1X_1T_{\rm h}>0$.
Then the heat flux into the hot heat reservoir
$\dot{Q}_{\rm h}\equiv P+\dot{Q}_{\rm c}\equiv J_3$ is given as
\begin{eqnarray}
J_3=P+\dot{Q}_{\rm c}=J_1X_1T_{\rm h}+J_2.\label{eq.minimal.J3}
\end{eqnarray}
By solving eq.~(\ref{eq.minimal.J1}) with respect to $X_1$ and substituting it
into eqs.~(\ref{eq.minimal.J2}) and (\ref{eq.minimal.J3}),
we can rewrite $J_2$ and $J_3$ by using $J_1$ instead of $X_1$ as (see fig.~\ref{fig.1}):
\begin{eqnarray}
J_2=\frac{L_{21}}{L_{11}}J_1+L_{22}(1-q^2)X_2-\gamma_{\rm c} {J_1}^2,\label{eq.J2.by.J1}\\
J_3=\frac{L_{21}}{L_{11}}\frac{T_{\rm h}}{T_{\rm c}}J_1+L_{22}(1-q^2)X_2+\gamma_{\rm h} {J_1}^2,\label{eq.J3.by.J1}
\end{eqnarray}
where $q\equiv \frac{L_{12}}{\sqrt{L_{11}L_{22}}}$ is the usual
coupling strength parameter ($|q| \le 1 $ from
eq.~(\ref{eq.onsager.coeffi.restrict})) and $\gamma_{\rm h}$ denotes 
the strength of the power dissipation into the hot heat reservoir as
\begin{eqnarray}
\gamma_{\rm h} \equiv \frac{T_{\rm h}}{L_{11}}-\gamma_{\rm c}>0.\label{eq.gh}
\end{eqnarray}

Eqs.~(\ref{eq.J2.by.J1}) and~(\ref{eq.J3.by.J1}) allow a clearer
description of the refrigerator than eqs.
(\ref{eq.minimal.J1}) and (\ref{eq.minimal.J2}) by considering $J_1$
as the control parameter instead of $X_1$ at constant $X_2$. So,
each term in eqs.~(\ref{eq.J2.by.J1}) and (\ref{eq.J3.by.J1}) has
explicit physical meanings (see also~\cite{yuki12}): the two first
terms mean the reversible heat transport between the working system
and the heat reservoirs; the second ones account for coupling
effects between the heat reservoirs; and the third ones account for the power dissipation ($\sim
J_1^2$) into the heat reservoirs, which inevitably occurs in a finite-time motion ($J_1 \ne 0$).

Using eqs.~(\ref{eq.J2.by.J1}) and (\ref{eq.J3.by.J1}) instead of eqs.~(\ref{eq.minimal.J1})
and (\ref{eq.minimal.J2}), the power input $P=J_3-J_2$ and the entropy production rate 
$\dot{\sigma}=\frac{J_3}{T_{\rm h}}-\frac{J_2}{T_{\rm c}}$ are written as
\begin{eqnarray}
P=\frac{L_{21}}{L_{11}}\left(\frac{1}{\tau}-1\right) J_1+\frac{T_{\rm h}}{L_{11}}{J_1}^2,\label{eq.P}
\end{eqnarray}
and
\begin{eqnarray}
\dot{\sigma}=L_{22}(1-q^2)X_2^2+J_1^2\left(\frac{\gamma_{\rm c}}{T_{\rm c}}+\frac{\gamma_{\rm h}}{T_{\rm{h}}}\right).
\end{eqnarray}

For the refrigerator, the converter efficiency $z$ in eq.~(\ref{eq.chi.def}) is the COP
$\varepsilon$ defined as
\begin{eqnarray}
\varepsilon\equiv \frac{\dot{Q}_{\rm c}}{P}=\frac{J_2}{P}=\frac{\frac{L_{21}}{L_{11}}J_1+L_{22}(1-q^2)X_2-\gamma_{\rm c}J_1^2}{\frac{L_{21}}{L_{11}}(\frac{1}{\tau}-1)J_1+\frac{T_{\rm h}}{L_{11}}J_1^2},\label{eq.cop}
\end{eqnarray}
where we used eqs.~(\ref{eq.J2.by.J1}) and (\ref{eq.P}).
We also introduce the $\chi$-criterion as a product of the COP $\varepsilon$ and the cooling power $J_2$,
identifying the cooling power $Q_{\rm in}/t_{\rm cycle}$ in eq.~(\ref{eq.chi.def}) 
as $\dot{Q}_{\rm c}=J_2$: 
\begin{eqnarray}
\chi=\frac{{J_2}^2}{P}=\frac{\left(\frac{L_{21}}{L_{11}}J_1+L_{22}(1-q^2)X_2-\gamma_{\rm
c}{J_1}^2\right)^2}
{\frac{L_{21}}{L_{11}}\left(\frac{1}{\tau}-1\right)J_1+\frac{T_{\rm h}}{L_{11}} {J_1}^2},\label{eq.chi}
\end{eqnarray}
where we used eqs.~(\ref{eq.J2.by.J1}) and (\ref{eq.P}). 
Though eq.~(\ref{eq.chi.def}) using $Q_{\rm in}/t_{\rm cycle}$ appears to be valid only for cyclic heat 
energy converters, eq.~(\ref{eq.chi}) using $J_2$ is valid even for steady-state refrigerators.
In order to get a positive cooling power $J_2>0$, 
we see from eq.~(\ref{eq.J2.by.J1}) that $J_1$ should be located in a finite range as
\begin{eqnarray}
&&\frac{L_{21}-\sqrt{D}}{2L_{11}\gamma_{\rm c}}<J_1< \frac{L_{21}+\sqrt{D}}{2L_{11}\gamma_{\rm c}}\label{eq.J1_range_nontight}, 
\end{eqnarray}
where the discriminant $D$ is given by
\begin{eqnarray}
D\equiv L_{21}^2+4L_{11}^2L_{22}\gamma_{\rm c}(1-q^2)X_2>0.\label{eq.def.D}
\end{eqnarray}
We consider the optimization of the $\chi$-criterion with respect to $J_1$ in this range. 
In the following, we first analyze the case $|q|=1$ called the tight-coupling condition. 
Afterward we consider the nontight-coupling case $|q|\ne 1$.
\subsection{Tight-coupling case $|q|=1$}
In the tight-coupling case, eq.~(\ref{eq.J1_range_nontight}) becomes
\begin{eqnarray}
0< J_1 < \frac{L_{21}}{\gamma_{\rm c}L_{11}}.\label{eq.J1_range_tight}
\end{eqnarray} 
We find $J_1=J_1^{\rm max\chi}$ which gives the maximum of $\chi$ by solving the equation $\partial \chi/\partial J_1=0$.
Noticing that $J_1=L_{21}/(\gamma_cL_{11})$ 
as the solution of $J_2=0$ is always one of the solutions of  $\partial \chi/\partial J_1=0$, 
we can simplify that equation to the following quadratic one:
\begin{eqnarray}
J_1^2+\frac{3L_{21}}{2\varepsilon_{\rm C}T_{\rm h}}J_1-\frac{L_{21}^2}
{2\gamma_{\rm c}L_{11}\varepsilon_{\rm C}T_{\rm h}}=0.\label{eq.quadratic}
\end{eqnarray}
Solving this equation, we obtain the sole physically acceptable solution
which satisfies ${\partial}^2 \chi/\partial {J_1}^2<0$ and  
locates in the range given by eq.~(\ref{eq.J1_range_tight}) as
\begin{eqnarray}
J_1^{{\rm max\chi}}=\frac{2{L_{21}}}{3\gamma_{\rm
c}{L_{11}}+\gamma_{\rm c}{L_{11}} \sqrt{8\varepsilon_{\rm
C}\left(1+\frac{\gamma_{\rm h}}{\gamma_{\rm c}}\right)+9}}
.\label{eq.J1.optimum}
\end{eqnarray}
Substituting eq.~(\ref{eq.J1.optimum}) into eq.~(\ref{eq.cop}),
we can obtain the general expression of the COP under maximum $\chi$-condition
for the model of minimally nonlinear irreversible refrigerators with the tight-coupling condition:
\begin{eqnarray}
\varepsilon_{{\rm max\chi}}=\frac{\sqrt{9+8\varepsilon_{\rm C}\left(1+\frac{\gamma_{\rm h}}{\gamma_{\rm c}}\right)}-3}{2\left(1+\frac{\gamma_{\rm h}}{\gamma_{\rm c}}\right)}.\label{eq.chi.optimum}
\end{eqnarray}
It is a monotonic decreasing function of the dissipation ratio $\gamma_{\rm h}/\gamma_{\rm c}$. 
Considering that the assumption $\gamma_{\rm c}>0$ and $\gamma_{\rm c}<T_{\rm h}/L_{11}$ 
due to the non-negativity of $\gamma_{\rm h}$ as 
in eq.~(\ref{eq.gh}), 
the range of the dissipation ratio for the tight-coupling case is restricted to 
\begin{eqnarray}
0<\frac{\gamma_{\rm h}}{\gamma_{\rm c}}<\infty.\label{eq.dissipation_ratio_range_q_1}
\end{eqnarray}
Then we can obtain the lower and the upper bounds by considering
asymmetric dissipation limits $\gamma_{\rm h}/\gamma_{\rm c} \to \infty$
and $\gamma_{\rm h}/\gamma_{\rm c} \to 0$, respectively, as 
\begin{eqnarray}
\varepsilon_{\rm max\chi}^- \equiv 0 < \varepsilon_{{\rm max\chi}} < \frac{\sqrt{9+ 8\varepsilon_{\rm C}}-3}{2}\equiv \varepsilon_{\rm max\chi}^+,\label{eq.cop.bounds_2}
\end{eqnarray}
in full agreement with the results reported in~\cite{tu12d} and expressed in above eq.~(\ref{eq.cop.bounds}).
Here we stress that eq.~(\ref{eq.cop.bounds_2}) has been derived in a model-independent way and  
is valid for general refrigerators, regardless of the types of refrigerator (steady-state or cyclic).
It is also interesting to see from eq.~(\ref{eq.cop.bounds_2}) that the lower bound for 
$\varepsilon_{\rm max\chi}$
is $0$, which is in contrast to the case of the heat engines (see eq.~(21) in~\cite{yuki12}).
One may feel that this is a counterintuitive result at a first glance since the limit of the zero dissipation 
$\gamma_{\rm c}\to 0$ in the cold heat reservoir seems to be an advantageous condition for a refrigerator. 
This behavior can be understood as follows: from eq.~(\ref{eq.J1.optimum}), 
$J_1^{\rm max\chi}$ diverges as $\sim O(\gamma_{\rm c}^{-1/2})$. 
Then we can confirm $J_2^{\rm max\chi}\sim O(\gamma_{\rm c}^{-1/2})$ and $P_{\rm max\chi}\sim O(\gamma_{\rm c}^{-1})$ from
eqs.~(\ref{eq.J2.by.J1}) and (\ref{eq.P}), respectively, 
which leads to $\varepsilon_{\rm max\chi}=J_2^{\rm max\chi}/P_{\rm max\chi}\sim O(\gamma_{\rm c}^{1/2})\to 0$.
This is an anomalous situation where the refrigerator transfers 
the infinite amount of heat from the cold heat reservoir 
to the hot one, consuming the infinite amount of work. 
\subsection{Nontight-coupling case $|q|\ne 1$}
\begin{figure}
\begin{center}
\includegraphics[scale=1.2]{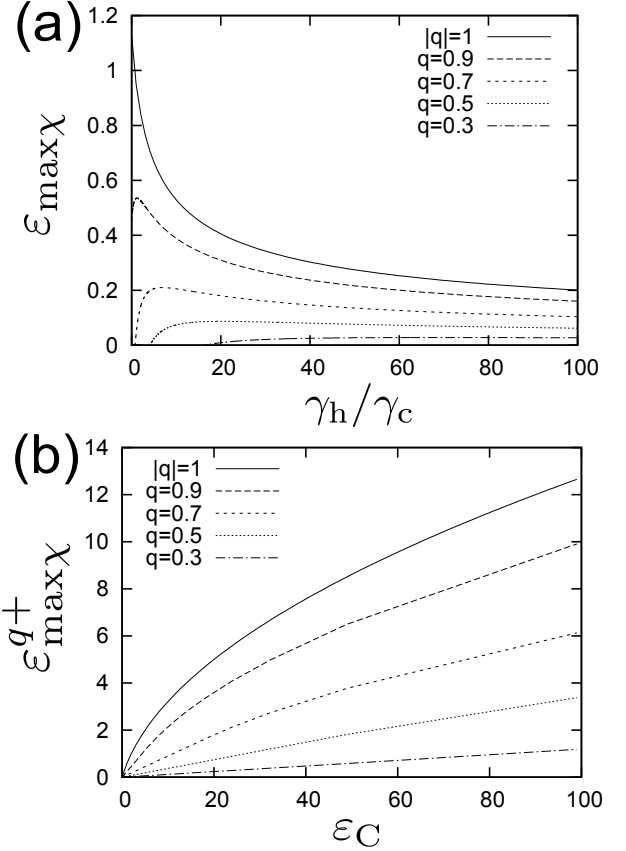}
\end{center}
\caption{Numerical solutions for the nontight-coupling case $|q|\ne 1$: (a) Dependence of  $\varepsilon_{\rm max\chi}$ 
on the dissipation ratio $\gamma_{\rm h}/\gamma_{\rm c}$ under given temperatures ($T_{\rm h}=1$ and $T_{\rm c}=0.7$) for some $q$'s. We also plot eq.~(\ref{eq.chi.optimum}) for the tight-coupling case $|q|=1$ for comparison.
We always set $L_{11}=1$, $L_{22}=1$. Giving $q$, we can determine $L_{21}$ 
as $L_{21}=\sqrt{q^2L_{11}L_{22}}$. 
(b) Dependence of the upper bound $\varepsilon_{\rm max\chi}^{q+}$ on  $\varepsilon_{\rm C}$  
for some $q$'s. We also plot the upper bound $\varepsilon_{\rm max\chi}^+$ for the tight-coupling case 
$|q|=1$ in eq.~(\ref{eq.cop.bounds_2}) for comparison.
We used the same values for $L_{ij}$'s as in (a).
We fixed $T_{\rm h}=1$ and changed $T_{\rm c}$.}\label{fig.2}
\end{figure}
Comparing with the tight-coupling case, 
we found that the analytic solution of the nontight-coupling case 
becomes too complicated to write down explicitly. 
Thus we here solve the optimization problem numerically. 
We first note that eq.~(\ref{eq.def.D}) implies the restriction
\begin{eqnarray}
\gamma_{\rm c}<-\frac{L_{12}^2}{4(1-q^2)L_{11}^2L_{22}X_2}\equiv \gamma_{\rm c}^+,\label{eq.gc_restrict} 
\end{eqnarray}
for given Onsager coefficients $L_{ij}$'s and $X_2$.
Such restriction does not appear in the tight-coupling case.
Besides, we note that we also have the restriction $0<\gamma_{\rm c}<T_{\rm h}/L_{11}$
due to the assumption of the non-negativity of $\gamma_{\rm c}$ and $\gamma_{\rm h}$ 
as in eq.~(\ref{eq.gh}). 
Thus, by combining these two inequalities, 
we obtain the restriction of the dissipation ratio for the nontight-coupling case,
depending on the value of $\gamma_{\rm c}^+$ as 
\begin{eqnarray}
0<\frac{\gamma_{\rm h}}{\gamma_{\rm c}}<\infty \label{eq.dissipation_ratio_range_q_ne_1_1}
\end{eqnarray}
in the case of ${T_{\rm h}}/{L_{11}}\le \gamma_{\rm c}^+ < \infty$, and 
\begin{eqnarray} 
\frac{\frac{T_{\rm h}}{L_{11}}-\gamma_{\rm c}^+}{\gamma_{\rm c}^+}<\frac{\gamma_{\rm h}}{\gamma_{\rm c}}<\infty
\label{eq.dissipation_ratio_range_q_ne_1_2} 
\end{eqnarray}
in the case of $0<\gamma_{\rm c}^+<{T_{\rm h}}/{L_{11}}$, respectively.

We can obtain $\varepsilon_{\rm max\chi}$ numerically as follows: 
we fix the values of $L_{11}$, $L_{22}$, $q$, $T_{\rm h}$ and $T_{\rm c}$. 
Choosing $\gamma_{\rm c}$ so as to satisfy $0<\gamma_{\rm c}<T_{\rm h}/L_{11}$ and eq.~(\ref{eq.gc_restrict}), 
which also determines $\gamma_{\rm h}$ from eq.~(\ref{eq.gh}),
we can find $\varepsilon$ at the maximum value of $\chi$ by calculating eq.~(\ref{eq.chi}) 
as $J_1$ is varied in the range of eq.~(\ref{eq.J1_range_nontight}). 

We first consider typical dependence of $\varepsilon_{\rm max\chi}$ 
on the dissipation ratio $\gamma_{\rm h}/\gamma_{\rm c}$ under given temperatures for some $q$'s.
By changing $\gamma_{\rm c}$, 
we plotted $\varepsilon_{\rm max\chi}$ as a function of $\gamma_{\rm h}/\gamma_{\rm c}$ in fig.~\ref{fig.2} (a).
As one of the interesting characteristics, 
we can see that the dependence can become non-monotonic and $\varepsilon_{\rm max\chi}$ 
attains the maximum value at a finite dissipation ratio for each $q$.
This contrasts to 
the tight-coupling case in eq.~(\ref{eq.cop.bounds_2}), 
where its upper bound $\varepsilon_{\rm max\chi}^+$ 
is always attained in the asymmetric dissipation limit $\gamma_{\rm h}/\gamma_{\rm c}\to 0$.

Next we determine the upper bound $\varepsilon_{\rm max\chi}^{q+}$ and 
the lower bound $\varepsilon_{\rm max\chi}^{q-}$ for the nontight-coupling case under various temperatures. 
By changing $\varepsilon_{\rm C}$, we evaluated the maximum value of $\varepsilon_{\rm max\chi}$
as $\varepsilon_{\rm max\chi}^{q+}$ for some $q$'s  (see fig.~\ref{fig.2} (b)).
We found that $\varepsilon_{\rm max\chi}^{q+}$ is always lower than $\varepsilon_{\rm max\chi}^{+}$ in eq.~(\ref{eq.cop.bounds_2}) 
and monotonically decreases as $q$ is decreased for each $\varepsilon_{\rm C}$. 
Meanwhile we also found that the lower bound $\varepsilon_{\rm max\chi}^{q-}$ agrees with 
$0$ for each $q$ (data not shown), which can always be attained in the asymmetric dissipation limit $\gamma_{\rm h}/\gamma_{\rm c} \to \infty$ 
as in the tight-coupling case in eq.~(\ref{eq.cop.bounds_2}). 
We also note that in the case where $\gamma_{\rm h}/\gamma_{\rm c}$ satisfies eq.~(\ref{eq.dissipation_ratio_range_q_ne_1_2}), 
the lower bound $\varepsilon_{\rm max\chi}^{q-}=0$ can be realized also at the lower endpoint of 
eq.~(\ref{eq.dissipation_ratio_range_q_ne_1_2}),
that is, in the limit of 
$\gamma_{\rm h}/\gamma_{\rm c}\to (T_{\rm h}/L_{11}-\gamma_{\rm c}^+)/\gamma_{\rm c}^+$.
(See the lines of $q=0.3$, $0.5$ and $0.7$ in fig.~\ref{fig.2} (a).)
In this case, 
$\varepsilon_{\rm max\chi}^{q-}$ becomes $0$ since $J_2^{\rm max\chi}$ vanishes: 
in the lower endpoint of eq.~(\ref{eq.dissipation_ratio_range_q_ne_1_2}),
the discriminant $D$ in eq.~(\ref{eq.def.D}) becomes $0$. Thus the solution of the inequality 
$J_2>0$ in eq.~(\ref{eq.J1_range_nontight}) approaches this multiple root in that limit, where $J_2$ 
in eq.~(\ref{eq.J2.by.J1}) is identically $0$.
This contrasts to the case of $\gamma_{\rm h}/\gamma_{\rm c}\to \infty$, where $J_2^{\rm max\chi}$ diverges.

\section{Example: leaky low-dissipation Carnot refrigerator}
As an example of the minimally nonlinear irreversible refrigerator, 
we here introduce a leaky low-dissipation Carnot refrigerator model.
The original low-dissipation Carnot refrigerator without heat leak 
studied in~\cite{carla12,tu12d} is an
extension of the quasistatic Carnot refrigerator by assuming that
heat transfer accompanying finite-time operation in each isothermal process is inversely proportional 
to the duration of the process $t_i$:
\begin{eqnarray}
Q_{\rm h}=T_{\rm h} \Delta S+\frac{T_{\rm h} \Sigma_{\rm h}}{t_{\rm h}},\label{eq.low.qh}\\
Q_{\rm c}=T_{\rm c} \Delta S-\frac{T_{\rm c} \Sigma_{\rm c}}{t_{\rm c}}, \label{eq.low.qc}
\end{eqnarray}
where we denote by $\Delta S>0$ the quasistatic entropy change of the working system during
each isothermal process, and by $\Sigma_i/t_i$ the corresponding
entropy production with a constant strength $\Sigma_i>0$~\cite{carla12,tu12d}.
Then we formally modify this low-dissipation Carnot refrigerator
to a leaky model by adding a heat leak term as follows:
\begin{eqnarray}
Q_{\rm h}=T_{\rm h} \Delta S+\frac{T_{\rm h} \Sigma_{\rm h}}{t_{\rm h}}-\kappa (T_{\rm h}-T_{\rm c})(t_{\rm h}+t_{\rm c}),\label{eq.low.qh.leak}\\
Q_{\rm c}=T_{\rm c} \Delta S-\frac{T_{\rm c} \Sigma_{\rm c}}{t_{\rm c}}-\kappa (T_{\rm h}-T_{\rm c})(t_{\rm h}+t_{\rm c}), \label{eq.low.qc.leak} 
\end{eqnarray}
where we assume the linear Fourier-type heat transfer from the hot heat reservoir to the cold one as the 
heat leak per cycle, denoting by $\kappa >0$ the effective thermal conductivity.
We here reinterpret $Q_{\rm h}$ ($Q_{\rm c}$) 
as the net heat transfer into (from) the heat reservoir per cycle, 
instead of the heat transfer in each isothermal process,
considering the heat leak contribution which always continues during one-cycle.

We show that this leaky low-dissipation Carnot refrigerator is an example of the 
minimally nonlinear irreversible refrigerator by rewriting eqs.~(\ref{eq.low.qh.leak}) 
and (\ref{eq.low.qc.leak}) into the forms of eqs.~(\ref{eq.J2.by.J1}) and (\ref{eq.J3.by.J1}):
since it works as a cyclic refrigerator and according to eqs.~(\ref{eq.def_J1_X1}) and (\ref{eq.def_J2_X2}), 
we can define its thermodynamic fluxes and
forces as $J_1\equiv 1/t_{\rm cycle}=1/(t_{\rm h}+t_{\rm c})=1/((\alpha+1)t_{\rm h})$,
$J_2\equiv \dot{Q}_{\rm c}=Q_{\rm c}/t_{\rm cycle}$, $X_1\equiv W/T_{\rm h}$ and $X_2\equiv
1/T_{\rm h}-1/T_{\rm c}$, with $\alpha \equiv t_{\rm
c}/t_{\rm h}$. By using these definitions and eqs.~(\ref{eq.low.qh.leak})
and (\ref{eq.low.qc.leak}), we can calculate the Onsager coefficients
$L_{ij}$'s and the strength of the dissipation $\gamma_i$'s as
\begin{eqnarray}
\left(\begin{array}{cc}
L_{11} & L_{12} \\ L_{21} & L_{22}
\end{array}\right)
=
\left(\begin{array}{cc}
\frac{T_{\rm h}}{Y}
&\frac{T_{\rm h} T_{\rm c} \Delta S}{Y} \\
\frac{T_{\rm h} T_{\rm c}\Delta S}{Y}
& \frac{{T_{\rm c}^2{T_{\rm h}}{\Delta S}^2}}{Y}+T_{\rm h}T_{\rm c}\kappa
\end{array}\right)
,\label{eq.onsager.coeffi}
\end{eqnarray}
\begin{eqnarray}
\gamma_{\rm h}=T_{\rm h}\Sigma_{\rm h}(\alpha+1),\label{eq.minimal.gammah}\\
\gamma_{\rm c}=\frac{T_{\rm c}\Sigma_{\rm c}(\alpha+1)}{\alpha},\label{eq.minimal.gammac}
\end{eqnarray}
respectively, where the common factor $Y$ in eq.~(\ref{eq.onsager.coeffi})
is given as
\begin{eqnarray}
Y\equiv (\alpha+1)\left(T_{\rm h}\Sigma_{\rm h}+\frac{T_{\rm c}\Sigma_{\rm c}}{\alpha}\right).
\end{eqnarray}
From eq.~(\ref{eq.onsager.coeffi}) we see that 
the Onsager reciprocity $L_{12}=L_{21}$ holds as expected.
In general, the Onsager coefficients show the nontight-coupling condition 
$|q|=\frac{L_{12}}{\sqrt{L_{11}L_{22}}} < 1$, but recover the tight-coupling condition $|q|\to 1$ 
in the zero leak limit of $\kappa \to 0$. 
Thus our model eqs.~(\ref{eq.J2.by.J1}) and (\ref{eq.J3.by.J1}) under the
tight-coupling condition includes the low-dissipation Carnot
refrigerator~\cite{carla12,tu12d} in eqs.~(\ref{eq.low.qh}) and (\ref{eq.low.qc}) as a special case.

Finally we note that we obtained eq.~(\ref{eq.chi.optimum}) for the tight-coupling case
maximizing $\chi$ only by one parameter $J_1$, whereas in the
original low-dissipation model without the heat leak
the optimization uses the two parameters $t_{\rm h}$
and $t_{\rm c}$~\cite{carla12,tu12d}. Thus the parameter space for the maximization is different. 
In the formalism presented here, a further optimization of 
$\chi=\chi(\alpha)$ by $\alpha$ is still possible 
after the optimization by $J_1$ and it corresponds to the two-parameter maximization done
in~\cite{carla12,tu12d}. 
But, that further optimization does not change the bounds in~eq.~(\ref{eq.cop.bounds}).
In particular, under symmetric conditions ($\Sigma_{\rm
h}=\Sigma_{\rm c}$) the further optimization by $\alpha$ gives $\alpha_{\rm
max\chi}=\sqrt{1-\tau}+1,\label{eq.alpha}$ from which we can
straightforwardly obtain the symmetric optimization COP in eq.~(3)~\cite{carla12}.
\section{Summary}
We introduced the extended Onsager relations as the
model of minimally nonlinear irreversible refrigerators, which can be applied to 
both steady-state and cyclic refrigerators. 
We analytically derived the general expression for the
coefficient of performance (COP) under maximum $\chi$-condition when the tight-coupling
condition is met and determined its upper and lower bounds which perfectly agree with the result derived previously in
the low-dissipation Carnot refrigerator model in~\cite{tu12d}.
We also studied the nontight-coupling case numerically and found that its upper bound is 
always lower than that of the tight-coupling case, whereas all the cases have the same lower bound.
Moreover, we introduce the leaky low-dissipation Carnot refrigerator.
Calculating the Onsager coefficients and the strength of the power
dissipation explicitly, we proved that this leaky low-dissipation Carnot refrigerator is an example of the minimally nonlinear irreversible refrigerator.

\acknowledgments
Y. Izumida acknowledges the financial support from a
Grant-in-Aid for JSPS Fellows (Grant No. 22-2109).
JMM Roco and A. Calvo Hern\'andez thank financial support from
Ministerio de Educaci\'on y Ciencia of Spain under Grant  No.
FIS2010-17147 FEDER.


\begin{thebibliography}{0}

\bibitem{esposito09}
  \Name{Esposito M., Lindenberg K. \and Van den Broeck C.}
  \REVIEW{Phys. Rev. Lett.}{102}{2009}{130602}.

\bibitem{benenti10}
  \Name{Benenti G., Saito K. \and Casati G.}
  \REVIEW{Phys. Rev. Lett.}{106}{2010}{060601}.

\bibitem{norma10}
  \Name{S\'anchez-Salas N., L\'opez-Palacios L., Velasco S. \and Calvo Hern\'andez A.}
  \REVIEW{Phys. Rev. E}{82}{2010}{051101}.

\bibitem{tu12b}
  \Name{Tu Z. C.}
  \REVIEW{Chin. Phys. B}{21}{2012}{020513}.

\bibitem{wu}
  \Name{Wu C., Chen L. \and Chen J.}
  \Book{Advances in Finite-Time Thermodynamics: Analysis and Optimization}
%  \Editor{A. Editor}
%  \Vol{9}
  \Publ{Nova Science, New York}
  \Year{2004}.
%  \Page{666}.

\bibitem{durmayaz}
  \Name{Durmayaz A., Sogut O. S., Sahin B. \and Yavuz H.}
  \REVIEW{Prog. Energy Combust. Sci.}{30}{2004}{175}.

\bibitem{tu08}
  \Name{Tu Z. C.}
  \REVIEW{J. Phys. A: Math. Theor.}{41}{2008}{312003}.

\bibitem{seifert07}
  \Name{Schmiedl T. \and Seifert U.}
  \REVIEW{Phys. Rev. Lett.}{98}{2007}{108301}.

\bibitem{seifert08}
  \Name{Schmiedl T. \and Seifert U.}
  \REVIEW{EPL}{81}{2008}{20003}.

\bibitem{seifert11}
  \Name{Seifert U.}
  \REVIEW{Phys. Rev. Lett.}{106}{2011}{020601}.

\bibitem{gaveau11}
  \Name{Gaveau B., Moreau M. \and Schulman L. S.}
  \REVIEW{Phys. Rev. Lett.}{105}{2011}{230602}.

\bibitem{esposito10a}
  \Name{Esposito M., Kawai R., Lindenberg K. \and Van den Broeck C.}
  \REVIEW{EPL}{89}{2010}{20003}.

\bibitem{esposito10b}
  \Name{Esposito M., Kawai R., Lindenberg K. \and Van den Broeck C.}
  \REVIEW{Phys. Rev. E}{81}{2010}{041106}.

\bibitem{abe11}
  \Name{Abe S.}
  \REVIEW{Phys. Rev. E}{83}{2011}{041117}.

\bibitem{blickle11}
  \Name{Blickle V. \and Bechinger C.}
  \REVIEW{Nature Phys.}{8}{2012}{143}.

\bibitem{curzon75}
  \Name{Curzon F. \and Ahlborn B.}
  \REVIEW{Am. J. Phys.}{43}{1975}{22}.

\bibitem{esposito10c}
  \Name{Esposito M., Kawai R., Lindenberg K. \and Van den Broeck C.}
  \REVIEW{Phys. Rev. Lett.}{105}{2010}{150603}.

\bibitem{yan12}
  \Name{Yan H. \and Guo H.}
  \REVIEW{Phys. Rev. E}{85}{2012}{011146}.

\bibitem{tu12a}
  \Name{Wang Y. \and Tu Z. C.}
  \REVIEW{Phys. Rev. E}{85}{2012}{011127}.

\bibitem{tu12c}
  \Name{Wang Y. \and Tu. Z. C.}
  \REVIEW{EPL}{98}{2012}{40001}.

\bibitem{carla12}
  \Name{de Tom\'as C., Calvo Hern\'andez A. \and Roco J. M. M.}
  \REVIEW{Phys. Rev. E}{85}{2012}{010104(R)}.

\bibitem{yan90}
  \Name{Yan Z. \and Chen J.}
  \REVIEW{J. Phys. D}{23}{1990}{136}.

\bibitem{velasco97a}
  \Name{Velasco S., Roco J. M. M., Medina A. \and Calvo Hern\'andez A.}
  \REVIEW{Phys. Rev. Lett.}{78}{1997}{3241}.

\bibitem{velasco97b}
  \Name{Velasco S., Roco J. M. M., Medina A. \and Calvo Hern\'andez A.}
  \REVIEW{Appl. Phys. Lett.}{71}{1997}{1130}.

\bibitem{allahv10}
  \Name{Allahverdyan A. E., Hovhannisyan K. \and Mahler G.}
  \REVIEW{Phys. Rev. E}{81}{2010}{051129}.

\bibitem{tu12d}
  \Name{Wang Y., Li M., Tu Z. C., Calvo Hern\'andez A. \and Roco J. M. M.}
   \REVIEW{Phys. Rev. E}{86}{2012}{011127}.

\bibitem{broeck05}
  \Name{Van den Broeck C.}
  \REVIEW{Phys. Rev. Lett.}{95}{2005}{190602}.

\bibitem{borja06}
  \Name{Jim\'enez de Cisneros B., Arias-Hern\'andez L. A. \and Calvo Hern\'andez A.}
  \REVIEW{Phys. Rev. E}{73}{2006}{057103}.

\bibitem{borja07}
  \Name{Jim\'enez de Cisneros B. \and Calvo Hern\'andez A.}
  \REVIEW{Phys. Rev. Lett.}{98}{2007}{130602}.

\bibitem{borja08}
  \Name{Jim\'enez de Cisneros B. \and Calvo Hern\'andez A.}
  \REVIEW{Phys. Rev. E}{77}{2008}{041127}.

\bibitem{yuki09}
  \Name{Izumida Y. \and Okuda K.}
  \REVIEW{Phys. Rev. E}{80}{2009}{021121}.

\bibitem{yuki10}
  \Name{Izumida Y. \and Okuda K.}
  \REVIEW{Eur. Phys. J. B.}{77}{2010}{499}.

\bibitem{yuki08}
  \Name{Izumida Y. \and Okuda K.}
  \REVIEW{EPL}{83}{2008}{60003}.

\bibitem{yuki12}
  \Name{Izumida Y. \and Okuda K.}
  \REVIEW{EPL}{97}{2012}{10004}.

\bibitem{O}
  \Name{Onsager L.}
  \REVIEW{Phys. Rev.}{37}{1931}{405}.

\bibitem{GM}
  \Name{de Groot S. R. \and Mazur P.}
  \Book{Non-Equilibrium Thermodynamics}
%  \Editor{A. Editor}
%  \Vol{9}
  \Publ{Dover, New York}
  \Year{1984}.
%  \Page{666}.

\bibitem{CW}
  \Name{Callen H. B. \and Welton T. A.}
  \REVIEW{Phys. Rev.}{83}{1951}{34}.

\bibitem{AOGGLa}
  \Name{Apertet Y., Ouerdane H., Glavatskaya O., Goupil C. \and Lecoeur Ph.}
  \REVIEW{EPL}{97}{2012}{28001}.

\bibitem{AOGGLb}
  \Name{Apertet Y., Ouerdane H., Goupil C. \and Lecoeur Ph.}
  \REVIEW{Phys. Rev. E}{85}{2012}{041144}.

\end{thebibliography}
\end{document}